\documentstyle[prl,aps,epsf]{revtex}
\begin{document}
\twocolumn[\hsize\textwidth\columnwidth\hsize\csname@twocolumnfalse%
\endcsname

\draft
%\preprint{SU-ITP \# 97/33}
\title{Recent Developments in the SO(5) Theory of High $T_c$ 
Superconductivity} 

\author{Shou-Cheng Zhang}
\address{\it Department of Physics, Stanford University, Stanford
  CA~~94305} 
\maketitle
\begin{abstract}
\end{abstract}
{\sl In this talk I outline the general strategy behind the
$SO(5)$ theory of high $T_c$ superconductivity. Progress in the 
direction of exact $SO(5)$ models, numerical exact diagonalization
and possible experimental tests are reviewed. I also address 
the criticisms raised recently against the $SO(5)$ theory and point out 
directions for future exploration. 
} 
]

In this brief review I would like to summarize the recent developments
in the $SO(5)$ theory of high $T_c$ superconductivity\cite{so5,demler}. 
Since my review in the Proceedings of the 
International Conference on Materials and Mechanisms
of Superconductivity\cite{m2s}, some important progress has been
made due to the efforts of many groups
\cite{meixner,junction,vortex,burgess1,rabello,henley,burgess2,proximity,eder}. 
There has also been the criticisms
raised. I shall start
by briefly outline the general idea and strategy behind this theory,
then proceed to discuss recent progress made in constructing exact
$SO(5)$ symmetrical models, evidence from exact numerical diagonalization
of the $t-J$ model (see Prof.
Werner Hanke's contribution to this Proceeding for more details),
and various proposals for experimental tests (see Prof.
John Berlinsky's contribution to this Proceeding for more details). 
I shall address some
criticisms raised recently by Greiter\cite{greiter,reply}, 
Baskaran and Anderson\cite{pwa},
both from model dependent and general considerations.

In many people's mind, the theoretical problem of high $T_c$ is to start 
from the ``realistic" model Hamiltonian, reduce it to a simple form and
solve it to see if it produces the generic phase diagram of the high $T_c$ 
superconductors. Ten years after the discovery of high $T_c$ superconductivity,
it becomes clear that this is strategy does not work very well. There are
no controlled ways of reducing a ``realistic" Hamiltonian to simpler forms
and even if one accepts simple models like the Hubbard or the $t-J$ model, one
can still not determine its phase diagram. In view of these difficulties,
we would like to propose a different strategy, essentially by reversing the
arrow of the above mentioned logic. {\it First let us start with the generic 
phase diagram of the cuperates, assuming that 
antiferromagnetism (AF) and d-wave superconducting (dSC)  order are
the only zero temperature
phases in the clean limit, and ask what kind of Hamiltonians can give this
type of phase diagram}. If we only require the model to have SC order in
its ground state, then the problem is trivial, since any model with purely
attractive interaction will surely do. However, such models will in general
not have AF order at half-filling. Models with AF order generally require
repulsive interaction, and a generic model will not give SC order 
away from half-filling. Therefore, the first problem we opposed
is highly non-trivial, but well-defined mathematically.
Supposed that this problem is solved, {\it our next goal is to show
that the special model Hamiltonians we found are adiabatically connected to the
real system}. This goal is less well defined, but can still be investigated
with a reasonable degree of satisfaction.  Because the AF and SC
phases are infrared fixed points in the RG sense and thus stable at zero temperature,
models which can produce both of them have a good chance of being adiabatically
connected. One can also investigate the special models and ``realistic" models
numerically, and try to follow the low energy levels as best as one can.
Having established the first two goals, {\it our final objective is to formulate
a general theory with a small number of phenomenological constants,
which captures all the qualitative physics of the special models and can be fitted
quantitatively to experiments}.

Notice that the above strategy is formulated in exact footsteps after the strategy
behind the Landau fermi liquid theory. I mention this theory only in the
context of its strategy, and I am not saying that it is applicable to the normal
state of high $T_c$ materials. In the Landau fermi liquid theory, the goal is to
understand real systems like metals, normal $^3H_e$ liquids, heavy fermion
compounds etc. The first goal is established by the free fermi gas model, which is
exactly solvable and is the only concrete example of a fermi liquid. The second
goal is established by Landau's celebrated adiabatic hypothesis. This hypothesis
has never been proven but is extremely powerful once it is accepted. It makes the
free fermi gas model a prototype model
for a wide class of systems which are far
from it in Hamiltonian space. The last objective is of course achieved by Landau's
phenomenological fermi liquid model which can offer quantitative analysis for many
experiments.

Given the historical success of Landau's fermi liquid theory, it appears that the
above formulated goals towards the high $T_c$ problem may indeed be realistic.
How can we then achieve the first step? This was a trivial step in the fermi
liquid theory but extremely difficult in the present context, for above mentioned
reasons. Since this is no trivial matter, we need a principle, rather than a 
trick, to solve the problem. 
In order to find models which can give {\it both} AF and dSC order, 
it is natural to build in the symmetry between these two phases. This is 
precisely the idea behind the $SO(5)$ theory. $SO(5)$ symmetry is a symmetry
between the AF and dSC phases, therefore, if models with exact $SO(5)$ symmetry have
AF order, it must also have dSC order. At this point, three independent groups
have constructed such {\it microscopic} $SO(5)$ models\cite{rabello,henley,burgess2}, 
and reference \cite{rabello} gave a general group theoretical 
classification for this class of models.
Some of these models may have neither AF nor dSC order, some of them have both,
but it is not possible to have one without the other. At half filling, the basic
reasons for the Hubbard model to have AF order also applies
to a large class of these models. Therefore, by the principle of $SO(5)$ symmetry,
we can state with the same degree of confidence that {\it we now have microscopic models 
which have both AF order at half-filling and dSC order away from half-filling.}
Important progress has also been made in the second stated goal. Exact diagonalization
studies on Hubbard and $t-J$ clusters\cite{meixner,eder} by Hanke's group
show striking evidence that the  
low energy states agree with the level structure anticipated from $SO(5)$ symmetry,
and therefore gives some indication that the microscopic models with exact $SO(5)$
symmetry can be adiabatically connected to more ``realistic" models. Progress
towards the final goal is made historically before the first two. The 
$SO(5)$ quantum rotor model with anisotropic couplings captures the basic
physics of the microscopic $SO(5)$ models, but contain phenomenological 
parameters which may eventually be determined 
experimentally\cite{vortex,proximity}. It is tempting to believe that
all models with this type of phase diagram flow to this theory in
the long wave length limit\cite{fradkin}.

In the problem of high $T_c$ superconductivity, we are not dealing with a
new state of matter. Both the AF and dSC phases are well-characterized
by their respective broken symmetries. However, while these two basic 
phases are known and well-understood, their intimate relationship 
is the key challenge in the high $T_c$ problem. Never before in 
condensed matter physics have we encounter the competition between
two diagonally opposite types of broken symmetries ---
namely the insulating diagonal-long-range-order
(DLRO) on one side and the superconducting off-diagonal-long-range-order
(ODLRO) on the other --- manifested on
such a grand scale. Because our understanding of
their competition and possible unification is of such a paramount
importance to condensed matter physics in general, one should be
given the license strip away all the model dependent details and
extract general principles which captures the core physics unifying 
AF and dSC. Unification of fundamentally different {\it forces} by symmetry
principle is a central scheme of the 20th century particle physics. The idea
behind the $SO(5)$ theory is that fundamentally different {\it phases}
in condensed matter physics can also be unified by similar symmetry
principles. 

After outlining the general strategy, let me now turn to review 
the details of the recent progress. Let me first start with the
microscopic $SO(5)$ models. This problem was solved 
by three independent groups\cite{rabello,henley,burgess2}. Two technical
innovations helped this progress. One is the Henely-Kohno factor
${\rm sgn}(\cos p_x - \cos p_y)$, 
introduced by Chris Henley and Hiroshi Kohno independently.
This factor has the symmetry property of $d-$wave, and its square is
unity. If one uses this factor in the definitions of the $\pi$
operators, rather than the conventional $\cos p_x - \cos p_y$ factor used by
Demler and myself\cite{demler,so5}, the $SO(5)$ algebra closes exactly. 
The second one is the concept of a spinor\cite{rabello} which
transforms according to the irreducible representation
of $SO(5)$. Therefore, one can construct general
exact $SO(5)$ symmetric models simply by enumerating the possible
bilinears that can be constructed from this spinor. The result is a
surprisingly large parameter space (or actually functions), which 
support exact $SO(5)$ symmetry. These models have the following 
attractive properties: 1) One can prove that if the model has AF
order at half-filling, it must have dSC order away from half-filling.
This fact follows from the $SO(5)$ symmetry and the fact that the
chemical potential term commutes with the Casimir operator of $SO(5)$.
2) The $\pi$ operators are {\it exact} eigen-operators of the Hamiltonian, and 
the $\pi$ resonance below $T_c$ predicted by Demler and myself\cite{demler}
is a {\it exact} excitation of the system. It has exactly the same quantum numbers,
the energy and intensity dependence on doping as the experimentally observed resonant
neutron scattering peaks in $YBCO$. 3) The SDW quasi-particles are related
to the BCS quasi-particles by exact $SO(5)$ rotations and 
the AF/dSC transition
is simply a ``gap rotation" transition, rather than a ``gap closing"
transition. 
This behavior may offer a basic explanation of the pseudo-gap behavior
observed in the underdoped cuperates, and can be directly tested
experimentally by construction suitable AF/dSC 
junctions\cite{junction,proximity} 
and by studying
the nature of the fermionic excitations at the boundary. In fact,
the $SO(5)$ generalized Bogoliubov-de Gennes equations should be
intensively studied.

Important progress has also been made in our second stated goal. Since the
microscopic $SO(5)$ models have the same phases as the real system, it is
quite plausible that they are in fact adiabatically connected. (It certainly
is more plausible than the fermi liquid hypothesis since broken symmetry phases
are much more robust than the fermi liquid states). However,
it is not so clear if they are actually adiabatically connected to the
models of the real system, since their phase diagrams have not yet
been established. We can try to establish the connection by
testing the low lying energy level
structures of the Hubbard or the $t-J$ model 
{\it numerically}\cite{meixner,rabello}. 
The $SO(5)$ symmetry predicts a well-defined structure of the
low energy states. For example, one can start with the one magnon state
at half-filling and analytically construct a $SO(5)$ rotated state 
by acting the $\pi$-annihilation operator on it. This state has two holes and
one can ask about its overlap with the ground state in the
two hole sector. If the ideas of $SO(5)$ work, this overlap should remain
finite in the infinite system size limit. Numerical results finds
clearly identifiable spectral peaks where these states should
overlap.
This process can be carried to higher doping, and it continue to work until
doping concentration exceeding 25\%. The energy splitting between the 
one magnon state at half-filling and the two hole ground state is a 
important parameter. It certainly depends on the chemical potential since these
two states have different electron number. However, once this number is fixed,
it should be the same number that determine splitting of other states connected
by the $\pi$ operator, e.g. the two magnon states, the triplet 2 hole state
and the 4 hole ground state etc. Therefore, the variance of the splitting
among various states connected by the $\pi$ operator is a well-defined numerical 
measure of how good is the $\pi$ operator a eigen-operator of the
$t-J$ model. For the $J/t=0.5$ model, this variance is about $0.07t$, 
much smaller compared to $J$.   

These numerical data are obtained from finite sized systems, and one should always
be careful with their interpretations. Usually, the energies do not change significantly
with system size, but the overlaps do. It is therefore important to check the
size dependence of the overlaps. But the near equality of the $SO(5)$ 
multiplet levels at the critical chemical potntial 
shed significant light on the nature of the transition between AF and dSC. 
Generically, the transition from AF to dSC is expected to be first order,
which could terminate at a critical point. But there is always a 
question on whether it is more Ising like with disjoint configurations
at both sides of the transition, or more like a Heisenberg-Neel model
in a field, where the order parameter rotates through a continuously
connected set of states with small energy barriers.
From the
$SO(5)$ multiplet structure, we can see how the superspin vector is rotated from
AF to dSC direction, and identify the intermediate states which connect them 
continuously and show that at the critical chemical potential, the energy barrier 
between AF and dSC is smaller than the natural parameters in the model, namely $t$ 
or $J$. This finding is extremely important, and I believe that it will not
change significantly with system size. It gives us confidence that the first 
order transition between AF and dSC is more Heisenberg like rather than Ising
like and is accompanied
by the soft collective excitations. These soft collective excitations may make the
dominant thermodynamic contributions in the underdoped regime and
may be responsible for the pseudogap physics.

Clearly, the microscopic $SO(5)$ models are only {\it prototype models}
for illustrating general principles and qualitative features, not to be 
used for quantitative
comparisons with experiments. On the other hand, 
the phenomenological $SO(5)$ quantum rotor
model\cite{so5,m2s} captures the basic physics, offers a simple and 
intuitive picture
of the AF/dSC transition in analogy with spin flop transition, and can 
in principle be used for quantitative comparison with experiments
and possibly predicting new effects.
The $SO(5)$ quantum rotor model is a bit like a Landau-Ginzburg theory
for the high $T_c$ problem. However, it goes much beyond the traditional
LG theory since it contains {\it quantum dynamics}.
The validity of this model to the real system rests on the principle of
adiabatic continuity. It can contain large $SO(5)$ anisotropies, as long
as the various collective modes remain the lowest energy excitations
within their respective quantum number sectors.
Recent developments in this area are the following: 
1) Using this type of models,
it was predicted that a superconducting vortex in the underdoped materials
has a AF rather than normal core\cite{vortex}. 2) It was used to explain the puzzling
long ranged proximity effect in $Pr$ doped $YBCO$ superconductors, and a
novel transition in the SC/AF/SC is predicted as a function of the AF
layer thickness or the applied current\cite{proximity}. 
3) The stripe phase observed in some
high $T_c$ materials may actually be a ``$SO(5)$ superspin spiral"\cite{spiral}. 
In this configuration, the superspin vector points in the AF direction on a 2-legged
ladder, in the dSC direction on the next 2-legged ladder, in the $\pi$
phases shifted AF direction on the next 2-legged ladder, and finally, in the
$\pi$ phase shifted dSC direction on the last 2-legged ladder, before the
structure repeats itself. The last statement is a new prediction. Therefore,
travelling in the direction transverse to the stripes, the $SO(5)$ superspin
spirals on a great circle. Prof. John Berlinsky will
summarize the details of these works in his contribution to the 
proceeding.

Recently, Greiter\cite{greiter}, Baskaran and Anderson\cite{pwa}
raised criticism against the $SO(5)$ theory.
Part of their comments are model dependent questions concerning the applicability of the
$SO(5)$ model to realistic systems and part Baskaran and Anderson's comment challenges the
core concept and the
overall direction of the $SO(5)$ approach. I shall briefly address the model dependent
part first, more details will be given by Prof. Hanke in his contribution. 
Greiter argues that the energy of the $\pi$ resonance {\it in the metallic phase}
is of the order of $U$, not $J$. This is clearly incorrect. The source of the
error has been traced in weak coupling\cite{reply}, where his initial arguments was based.
Here we present the strong coupling version of the argument.
The $\pi$ operator is
a spin triplet operator, therefore, the mutual interaction among the two electrons
inside the pair can only be of the order of $J$. However, when the $\pi$ pair is
injected into a metallic (or superconducting) state, its energy can be of the order
of $J$, $U$ or $2U$, depending on 
whether the $\pi$ pair goes into two empty sites,
one empty and one singly occupied sites or two singly occupied sites. Therefore,
the $\pi$ spectra should have
three peak structure, with the two high energy
peaks smeared into bands due to scattering. As long as the system is less than half-filled,
the low energy peak will have a finite spectral weight, proportional to doping $x$.
Numerical results on the Hubbard model
clearly demonstrate this peak structure and showing that {\it
the low energy peak scales inversely
with $U$}. For low energy physics, we are only interested in the lowest peak. 
The difference
between the spectral distribution and the average spectral energy could be the 
source of the confusion. While Greiter's arguments are incorrect for the
metallic or the superconducting state, they are applicable to the {\it insulating
state} where the chemical potential is discontinuous, and there are no
empty sites. The implication of the chemical potential discontinuity 
was already worked out in reference \cite{so5}.
The $\pi$ doublet excitation of the insulator splits in energy when the
chemical potential is varied from the center of the gap.
At the critical
chemical potential $\mu_c$, the energy of the $\pi^-$ mode vanishes and
that of the $\pi^+$ mode remains finite. This energy difference reflects
the discontinuity of the chemical potential at half-filling.

The beginning part of Baskaran and Anderson's comment addresses the stability of the
$\pi$ mode against possible perturbations. The effect of the next
nearst neighbor $t'$ is
a important issue and is not fully understood analytically, but
numerical calculations do show that the $\pi$ mode is stable
against $t'$. The problem is that one 
obtains a $t'$ contribution from band calculations which fits the shape of the $YBCO$
fermi surface but grossly over-estimates the band dispersion around the $(\pi,0)$
and $(0,\pi)$ points by at least one order of magnitude. Both photoemission experiments
and numerical diagonalizations show that the band dispersion around
these points 
is less than $10meV$. The $\pi$ mode is essentially formed by the multiple scattering
around these points and it can only be stable against $t'$ if the band dispersion 
at these points are small. It is very hard for current analytical calculations to produce 
this band narrowing effect, since it is of a subtle many-body origin,
and check the stability of the $\pi$ mode. However, numerical
calculations on the Hubbard and $t-J$ model show that the $\pi$ resonance is actually
stable against $t'$. The next question raised by Baskaran and Anderson concerns the
effect of $V$, the next nearst neighbor interaction. Their question about the 
$-1/4 n_i n_j$ term in the $J$ part of the Hamiltonian is mathematically identical to
the question of a general $V$ term. Unlike the Hubbard interaction 
which only acts on the same
site and has no effect on the mutual interaction within the $\pi$ pair, the $V$ term 
certainly shifts the energy of the $\pi$ peak. However, it also shifts the energy of 
a $d-$ wave hole pair, in very much the same way. Therefore, one would expect on general
grounds that $V$ does not affect the difference between the energy of a $\pi$ pair
and a $d-$ wave hole pair. Numerically, this indeed seems to be the case. 
In the actual experiment, the observed neutron resonance corresponds to
a process where a $d-$ wave hole pair is extracted and a $\pi$ pair inserted. {\it The 
resonant energy therefore measures the difference in their energies}.  
Computer calculations can miss small energy differences due to finite size effects, but a 
effect on the order of $eV$, as argued by Baskaran and Anderson, can certainly be
distinguished. 

These types of model dependent debates are important at a later stage 
of the development in any theories on high $T_c$, when one compares 
quantitative predictions with experiments.
At the current early stage of the $SO(5)$ theory, 
it is much more important to make sure
that the basic core ideas are correct and not in conflict with 
well-established general principles. That is why the general criticism 
raised by Baskaran and Anderson
should be the main focus of the debate. {\it Neither the $SO(5)$ theory nor 
the RVB theory are in conflict with any fundamental principles of physics,
the difference between them lies in the strategies of attacking the high
$T_c$ problem}. In their comment, Baskaran and Anderson quoted the
``Elitzur's theorem", and gave the impression that it is a fundamental
and rigorous result which is in conflict with the basis of the $SO(5)$
theory. In fact it is a result that has no direct applicability to
most condensed matter physics models in consideration. The ``Elitzur's
theorem" states that local gauge symmetry can not be broken spontaneously.
In condensed matter physics, the only local gauge symmetry is the
freedom of choosing the phase of a wave function locally, at the expense
of a gauge transformation on the electromagnetic vector potential. The
consequence of the Elitzur's theorem is to give the phase mode of a 
superconductor a finite mass, which is the well-known Anderson-Higgs
mechanism. In any model without the real electromagnetic fields, like
the Hubbard or the $t-J$ model, there is no {\it physical}
local gauge symmetry at all. However, local gauge symmetry can be
introduced artificially if one enlarge the Hilbert space artificially,
for example by representing the Heisenberg spin operator in terms
of bosons or fermions.
In this case, the manifestation of the ``Elitzur's theorem" is nothing
but the projection back
into the actual physical space itself! This kind of artificial
enlargement of the Hilbert space maybe useful in some approximate
mean field theory approaches, but are not suitable  
in discussing matters of fundamental principles.

Therefore, the true debate should be focused on the overall strategy and
philosophy towards attacking the high $T_c$ problem. The basic 
difference lies in the starting points. From the insulating side, is
it better to start from a AF state or a Mott insulator?
From the metallic side, is it better to start from the actual
SC ground state or a ``Luttinger liquid" state?
The first question of course is only well-defined if the Mott 
insulator state is a RVB spin liquid state without any long range
order. However, soon after the discovery of high $T_c$ superconductivity,
it was established both experimentally and theoretically that the
Heisenberg model on a square lattice does have long range order. 
On frustrated lattices, spin Hamiltonians either find other types
of order or go into spin-Peirls states. It appears that the 
RVB type of order could only occur at the quantum critical points
between these various types of order, 
and it may not exist as a phase in two or higher dimensions. 
Sometimes, the dichotomy at half-filling 
is formulated by contrasting the
spin-density-wave (SDW) picture to the Heisenberg-Neel picutre of the
AF state. There is however no qualitative difference
between these two, since they are characterized by the same type
of broken symmetry. Furthermore, Schrieffer, Wen and I\cite{swz} have
shown that these two extremes can be smoothly connected 
{\it quantitatively}, and the SDW approach can capture the 
physics at strong coupling as well. From the point of view of 
adiabatic continuity and symmetries, it is perfectly OK to think of 
the AF state as the result of a fermi surface instability, 
in the same way that the SC state is the result of another 
type of fermi surface instability. The difference in the size of their
gaps is only a quantitative matter. The dichotomy between the
AF state and the RVB state at half-filling is translated into the
dichotomy between the dSC state and the ``Luttinger liquid" state
away from half-filling. Once again, systematic calculations 
failed to show Luttinger liquid behavior in two dimensions. The only
instability of the fermi liquid away from commensurate fillings appear
to be SC order. Experimentally, the zero temperature state at optimal
doping is a dSC state rather than a Luttinger liquid state.
Therefore, like the RVB state, the Luttinger liquid
may not be a phase but only a quantum critical point between phases. 

Laughlin\cite{bob} has recently proposed that the main conceptual differences
between the RVB type and the $SO(5)$ type of theories can be removed
if one accepts the idea that RVB, Luttinger liquid states and
spin-charge separation only
exist as isolated points in the Hamiltonian space in two dimensions,
therefore, it may not be in conflict with the general ideas of
a quantum critical point\cite{qcp} and also not in conflict with the $SO(5)$
idea in particular, since 
it is well-known that {\it extra symmetry can be present at critical
points}. 
These points only exist because of the competition and unification
between various types of order. What are the most important and 
robust types of order in the high $T_c$ problem? Obviously they
are the AF and dSC order. From this point of view, RVB and $SO(5)$
theories are actually addressing the same type of physics. Laughlin's
insight not only offers a possible unification of two seeming 
divergent theoretical ideas, but may also lead to practical progress
by complimenting the ordering physics near the AF/dSC transition
with the novel transport physics in the high $T_c$ superconductors. 

In conclusion I summarized here recent results in the $SO(5)$ theory
of high $T_c$ superconductivity. Rapid progress has been made towards
the three stated goals of this approach. The model depedent part of
the criticism has been answered both by general arguments and detailed
numerical calculations. The difference between the $SO(5)$ theory and the
RVB theory mainly lies in the overall strategy of attacking the high $T_c$ 
problem, and only time could tell which one would work better.
Major direction for future theoretical development should concentrate
on the possibility of zero temperature $SO(5)$ symmetric quantum
critical point, more quantitative numerical tests of the $SO(5)$
symmetry in known models, working out quantitative predictions from
the $SO(5)$ quantum rotor model and studying the nature of the
fermionic excitations in the AF/dSC transition region.
In comparison with experiments, it is most important to establish
the high $T_c$ phase diagram in the AF/dSC transition regime.
$SO(5)$ theory is a rather bold hypothesis, and many striking
qualitative experimental predictions remain to be worked out.
The strange and mysterious quantum mechanical world where DLRO and 
ODLRO are unified
must have profound experimental manifestations filled surprises and
puzzles.

I would first like to thank D. Arovas, Y. Bazaliy, 
J. Berlinsky, E. Demler, R. Eder, W. Hanke, S. Meixner,   
C. Kallin, H. Kohno, S. Rabello and D. Scalapino, results reported
here are obtained through our close collaborations.
I would like to thank Prof. Laughlin for many
insightful comments which helped us to formulate the strategy
of our approach. 
This work is supported by the NSF under grant numbers DMR-9400372 
and DMR-9522915. 

\hspace{1cm}

\end{document}